# Advanced Trace Pattern For Computer Intrusion Discovery

Siti Rahayu S., Robiah Y., Shahrin S., Mohd Zaki M., Faizal M.A., and Zaheera Z.A **Abstract**—The number of crime committed based on the malware intrusion is never ending as the number of malware variants is growing tremendously and the usage of internet is expanding globally. Malicious codes easily obtained and use as one of weapon to gain their objective illegally. Hence, in this research, diverse logs from different OSI layer are explored to identify the traces left on the attacker and victim logs in order to establish worm trace pattern to defending against the attack and help revealing true attacker or victim. For the purpose of this paper, it focused on malware intrusion and traditional worm namely sasser worm variants. The concept of trace pattern is created by fusing the attacker's and victim's perspective. Therefore, the objective of this paper is to propose a general worm trace pattern for attacker's, victim's and multi-step (attacker/victim)'s by combining both perspectives. These three proposed worm trace patterns can be extended into research areas in alert correlation and computer forensic investigation.

—————————— ◆ ——————————

## 1 INTRODUCTION

MALWARE has become a serious threat to the economy and to national security in recent years. Malware or malicious software is software that is residing in a system and it is intended to cause harm to the system. Malware that consist of Trojan, virus and worm had threatened the internet user and causes billion of losses to the internet users around the world. As computer users rely ever more on the Internet to use the online services, they face complex challenges in securing information systems and networks from attack or penetration by malicious software that anytime can steal their credential information.

In 2009 alone, according to anti-virus vendor Panda Security [1] 25 million new malware samples have been found luring on the Internet and this indicate that the level of the malware threat has tremendously increase and something has to be done to safeguard the Internet user from the threat. Malware especially worm is difficult to detect especially with its ability to change it behavior of infecting others system make the antivirus seem difficult to notice them. The variants of worm are often created to defeat the security tools, for instance a worm can mutate to a different variants, sometimes in only one hour [2]. Thus make it difficult for security tool to detect the threat.

As a result, the study on internet attack or intrusion is very crucial, especially in developing an effective security tool to defend the internet user from the attack threat. This phenomenon was create a relative common task for security researchers to collect data related to Internet threats to help the researchers to investigate the trace pattern in order to find the root cause and effect of an intrusion in victim and attacker perspectives based on the intrusion's anatomy described in [3]. Trace pattern can also be used as a guide to the investigator for collecting and tracing the evidence in forensic field [4].

To address this adversity, the attacker's, victim's and multi-step (attacker/victim)'s trace patterns is proposed in this paper by fusing both attacker and victim perspectives. The various logs from different OSI layer are explored in this research to identify the traces leave on the attacker and victim logs and establish the general trace pattern that used to reveal true attacker or victim. For the purpose of this paper, the research only focuses on malware network intrusion specifically on sasser worm variants in three different intrusion scenarios namely Scenario A, Scenario B and Scenario C.

## 2 RELATED WORK

### 2.1 Sasser Worm

Worm is a one of self-replicating Malware [1] that uses computer network to automatically send duplicate copies of itself to other vulnerable host connected to the network. In this paper, the researcher focuses only on traditional worm specifically blaster and sasser worms as described by [2].

Sasser was first noticed and started spreading on April 30th, 2004. It is a computer worm that affects computers running vulnerable versions of the Microsoft operating systems *Windows XP* and *Windows 2000*. This worm was named Sasser because it spreads by exploiting a buffer overflow in the component known as *LSASS* (*Local Security Authority Subsystem Service*) on the affected operating systems. Sasser is programmed to launch 128 processes which scan a range of random IP addresses looking for

- *Siti Rahayu Selamat is with the FTMK, Universiti Teknikal Malaysia Melaka.*
- *Robiah Yusof is with the FTMK, Universiti Teknikal Malaysia Melaka.*
- *Shahrin Sahib is with the FTMK, Universiti Teknikal Malaysia Melaka.*
- *Mohd Zaki Masu'd is with the FTMK, Universiti Teknikal Malaysia Melaka.*
- *Mohd Faizal Abdollah is with the FTMK, Universiti Teknikal Malaysia Melaka.*
- *Zaheera Zainal Abidin is with the FTMK, Universiti Teknikal Malaysia Melaka.*

systems vulnerable to the LSASS vulnerability on port *445/TCP*. Then, it installs an FTP server on port *5554* so that it can be downloaded by other infected computers. Once a vulnerable machine is found, the worm opens a remote shell on the machine (on *TCP port 9996*), and makes the remote machine download a copy of the worm namely *avserve.exe* or *avserve2.exe* for the Sasser.B variant in the Windows directory [5].

In order to find new victims, Sasser scans random IP addresses for vulnerable machines listening on port *445/TCP*. Once such a machine is found, it attempts to exploit the *LSASS* vulnerability by sending a specially crafted RPC request to the *LSASS* named pipe on the machine. Upon successful exploitation, shell code is injected into the *lsass.exe* process, which executes a shell (*cmd.exe*) and binds it to a TCP port. The attacking instance of the worm then connects to this port and sends commands to the shell. These commands download and run the main worm executable on the newly infected system. The worm download is carried out through FTP, using the default *Windows ftp.exe* program on the client side (victim). On the server side (attacker), Sasser implements its own crude FTP server, which listens on a non-standard TCP port [6]. In this research, the researchers had discovered the sasser variant used in the experiment using non-standard *TCP port 3\*\*\**.

The infection scheme of Sasser is very similar to that of W32/Blaster with the exception of using FTP instead of TFTP as the main transmission protocol. The scanner threads attempts to determine the local machine's IP address. It loops through every address returned by *gethostbyname* for the local hostname. If it finds a publicly routable Internet address (non-RFC1918) it will use that address. If none are found it will use any private subnet address (RFC1918 or 127.0.0.1) it finds. If no address is returned it will use 127.0.0.1. The way a target IP to exploit is generated is that; 50% of the time it will attempt to exploit a completely random IP address, 25% of the time it will attempt to exploit a random address within the same first octet of the local subnet and 25% of the time it will attempt to exploit a random address within the same first and second octets of the local subnet. The aim is to increase the probability of hitting vulnerable hosts, on the assumption that nearby machines suffer from the same misconfiguration problems. The network scanning speed per attack thread is a maximum of four attacks per second, and Sasser spawns 128 attack threads running in parallel [6].

If successful, the *LSASS* exploit will open a shell on the remote system on *TCP port 9996*. The worm will connect to this port and attempt to send the following commands:

> *echo off&echo open (infecting machine's IP)*
> *5554>>cmd.ftp&echo anonymous>>cmd.ftp&echo user&echo bin>>cmd.ftp&echo get*
> *(rand)_up.exe>>cmd.ftp&echo bye>>cmd.ftp&echo*
> *on&ftp -s:cmd.ftp&(rand)i_up.exe&echo off&del*
> *cmd.ftp&echo on*

This will copy the worm executable to the target machine, where it will run and begin to spread and the thread sleeps for 250 milliseconds, and then repeats the entire process again.

When executed, the worm will installs itself to *%WINDIR%* as *avserve.exe* and adds the following registry key *HKLM\Software\Microsoft\Windows\ CurrentVersion\Run. avserve.exe -> C:\%WINDIR%\avserve.exe*. It will creates a Mutex "Jobaka31" to ensure that only one copy of the worm runs in memory. Then it will spawn a mini-FTP server on *TCP port 5554* to deliver the worm executable to exploited systems. It will generate 128 threads to scan for and exploit vulnerable systems. After that it will calls API method AbortSystemShutdown to prevent the system from rebooting. It will then sleeps for 3 seconds and then loops back to the AbortSystemShutdown call.

An indication of the worm's infection of a given PC is the existence of the file *C:\WIN.LOG* or *C:\WIN2.LOG* on the computer hard disk, and random crashes of *LSASS.EXE* caused by faulty code used in the worm. The most common characteristic of the worm is the shutdown timer that appears due to the worm crashing LSASS.exe as described in [7].

**2.2 Trace Pattern**

Trace pattern is defined as a regular way of process discovering the origin or starting point of a scenario that has happened [8]. It is an essential element in helping investigator in a crime scene to find the evidence, for instance in a computer crime the evidence can be found in any digital devices. The evidences of a computer crime can be in form of data records that consists of user activities such as login, logout, computer shutdown, files execution and network packet. In typical digital devices all these traces data are presented on logs file, such as the data records in a network log files consist of several selected attributes such as port, action, protocol, source IP address and destination IP address.

In forensic view, a victim or attacker can be identified based on the traces data found in the attack pattern analysis and represent in the form of trace pattern in which the trace pattern can help determine how a crime is being committed. Attack pattern is type of pattern that is specified from attacker perspective. The pattern describes how an attack is performed, enumerates the security patterns that can be applied to defeat the attack, and describes how to trace the attack once it has occurred [9].

An attack pattern presents a logical description of the attack goals and attack approaches for defending against and tracing the attack. Hence, attack patterns can guide forensic investigators in searching the evidence and the patterns can serve as a structured method for obtaining and representing relevant network forensic information. This also helps the forensic investigator at the data collection phase that requires the investigator to determine and identifying all the components to be collected, deciding the priority of the data, finding the location of the com-ponents and collecting data from each of the component during the investigation process [10].

There are various explanations on describing the term attack pattern. In general, researches describe the term attack pattern as the steps in generating attack and exploiting the target as mentioned in [11], [12], [13] [10], [9] and imperative to provide a way to protect them from any potential attack. However, all of the researchers are only concentrating on the attacker's perspective without considering the victim's perspective. Therefore, the trace patterns are proposed in this research by deliberating on the attacker's, victim's and attacker/victim's (multi-step) perspectives in order to have a clear view on how the attack is performed and caused the impact to the target. In this research, trace pattern on multi-step perspective is established that is motivated based on the study by [14] to help the investigator on revealing the true attacker or victim in which [15] describes a multi-step attack is a sequence of attack steps from the attacker performed the attack until the compromised host start generating a new attack to another target.

In the next section, researchers present the intrusion scenario used in this research to gather and analyse logs for designing the proposed worm trace pattern.

## 3 INTRUSION SCENARIO

A controlled experiment is designed in this research in order to run the worm intrusion, to collect logs from each of the devices involved and to design the intrusion scenario. This experimental approach used four phases: *Network Environment Setup*, *Attack Activation*, *Trace Pattern Log Collection* and *Trace Pattern Log Analysis* as described in [8]

In this experiment, the worm intrusion is launched and the intrusion activities are captured in the selected logs which are *personal firewall log*, *security log*, *system log*, *application log*, *IDS log*, *tcpdump* and *Wireshark log*. The researchers have collected all logs generated during the experiment and three intrusion scenarios are derived based on the log analysis are identified as Scenario A, Scenario B and Scenario C as depicted in Fig. 1, Fig. 2 and Fig. 3 respectively. Each analysis for each intrusion scenario involved with selected logs that divided into host level: security log, application log, system log and personal firewall log and network level: *IDS Alert log*.

Based on the experiment setup, the researcher launched the attack host *Selamat* in all three scenarios identified. However, the victim and the victim/attacker for each scenario were different. In Scenario A as in Fig. 1, *Selamat* is successfully exploited host *Roslan* that mark with *445, 9996, 5554* and *3\*\*\**. However, host *Yusof* was mark with *445* and *9996* that describes the attacker is already open the backdoor but unable to transfer the malicious codes through port *5554*. Then, the compromised host (*Roslan*) has begun a new attack on host *Mohd*.

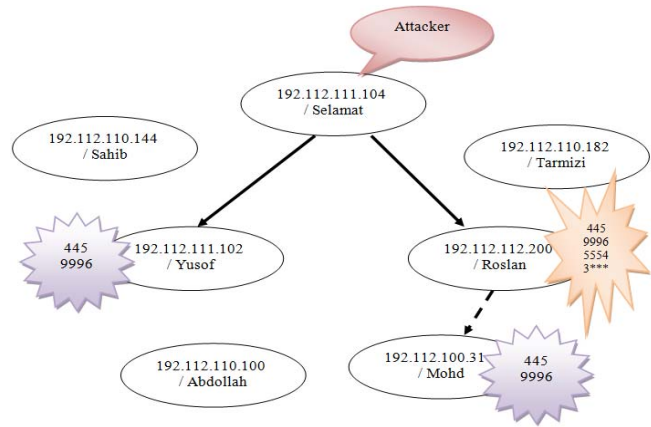

Fig. 1 Sasser Intrusion Scenario: Scenario A

On the other hand, Scenario B as in Fig. 2 demonstrates that host *Roslan* and *Ramly* became the targets of the attack. *Selamat* is completely exploited host Ramly that marks with *445, 9996, 5554* and *3\*\*\**, but unable to transfer the exploit codes to Ramly which indicates with *445* and *9996*. Subsequently, the infected host (*Ramly*) managed to exploit host Roslan completely (*445, 9996, 5554* and *3\*\*\**). Next, once the host *Roslan* is infected, it continually launches an attack and successfully exploit host *Sahib*.

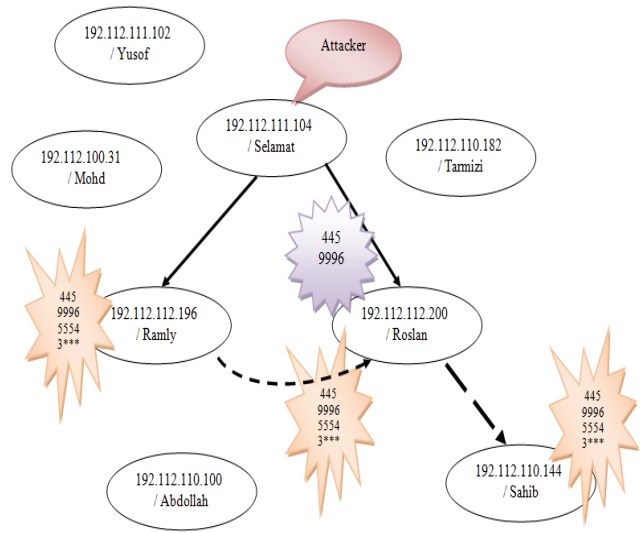

Fig. 2 Sasser Intrusion Scenario: Scenario B

Meanwhile in intrusion Scenario C as represented in Fig. 3, *Selamat* was successfully exploit host *Sahib* (*445, 9996, 5554, 3\*\*\**) and manage to open the backdoor at host Tarmizi (445, 9996) but unable to upload the exploit codes through port *5554*. Once *Sahib* is infected, it automatically generates an attack to another target which still uninfected. In this scenario, *Sahib* is successfully exploited *Tarmizi* since *Tarmizi* is still uninfected host.

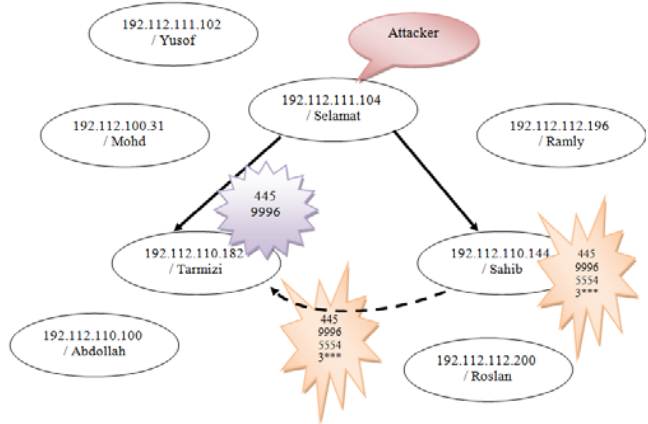

Fig. 3 Sasser Intrusion Scenario: Scenario C

The intrusion scenarios are summarized as in TABLE 1 in which the attack was launched from host *Selamat* (192.112.111.104).

TABLE 1
SUMMARY OF INTRUSION SCENARIO

| Intrusion Scenario | Origin of Attack | Exploited Node (Victim) | Compromised Node (Victim/Attacker) | |
|---|---|---|---|---|
| | | | Level 1 | Level 2 |
| Scenario A | 192.112.111.104 / Selamat | 192.112.111.102/ Yusof<br>192.112.112.200/ Roslan | 192.112.112.200/ Roslan | - |
| Scenario B | 192.112.111.104 / Selamat | 192.112.112.196/ Ramly<br>192.112.112.200 / Roslan | 192.112.112.196/ Ramly | 192.112.112.200 / Roslan |
| Scenario C | 192.112.111.104 / Selamat | 192.112.110.182 / Tarmizi<br>192.112.110.144/ Sahib | 192.112.110.144/ Sahib | - |

*Roslan*, *Ramly* and *Sahib* were successfully exploited by *Selamat* in Scenario A, Scenario B and Scenario C respectively and became an attacker to another uninfected host that shown as compromised node (victim/attacker): Level 1. In compromised node in Level 2, it describes that *Roslan* in Scenario B was attacked and exploited by *Ramly* which was infected previously in Level 1. These attacks are known as multi-step attack although *Roslan* and *Ramly* are became a victim or attacker in different level.

## 4 ANALYSIS AND FINDINGS

Three intrusion scenarios discussed in section III are further analyzed and the findings from this analysis are used as the primary guideline in establishing the generic worm trace pattern in victim, attacker and multi-step (victim/attacker) perspectives by observing the traces leave on the selected logs. The details of the trace pattern analysis from each perspective are explained in the following sub-section.

### 4.1 Worm Trace Pattern Analysis: Victim Perspective

The victim's data traces are discovered from the logs at the victim's host and network log. The summary of the data traces for all scenarios derived as discussed in section III are shown in TABLE 2 and the evidences are found in *personal firewall log*, *security log*, *system log*, *application log* and *alert IDS log* based on the traces present in each log that significant to the intrusion trace pattern.

In *Personal Firewall log* for each scenario, there are significant vulnerable ports exist that can be used by malicious codes to exploit which exploits a security hole in the *LSASS* (Local Security Authority Subsystem Service, which corresponds to the executable file lsass.exe) in Windows on its victims which are *TCP/445* that used for worm scanning activity and port *TCP/9996*, *TCP/5554* and *TCP/3\*\*\** that is used for exploiting activity. According to [16],[17], [17] and [5], this traces data is considered as part of this victim's trace. In this analysis, researchers found trace of port *TCP/3\*\*\** is used by sasser to transfer the exploit code and without this port is opened, the attacker unable to transfer the malicious codes to the target or victim although port *TCP/5554* is opened.

TABLE 2
SUMMARY OF SASSER TRACES ON VICTIM'S LOG FOR SCENARIO A, SCENARIO B AND SCENARIO C
(√=TRACES FOUND; x=TRACES NOT FOUND)

| Perspective | Level | Trace Category | Evidence /Log | Trace Pattern | Scenario | | | General Victim Trace Pattern | |
|---|---|---|---|---|---|---|---|---|---|
| | | | | | A | B | C | Trace Discovered | Trace Attributes |
| Victim | Host | Scan | Personal Firewall | 445 OPEN-INBOUND TCP | √ | √ | √ | √ | Action<br>Protocol<br>Destination Port |
| | | Exploit | Personal Firewall | 9996 OPEN-INBOUND TCP<br>5554 OPEN TCP<br>3\*\*\* OPEN-INBOUND TCP | √ | √ | √ | √ | Action<br>Protocol<br>Destination Port |
| | | Impact | Security | Event ID: 592<br>IFN: %WINDIR%\system32\ftp.exe<br><br>Event ID: 592<br>IFN: %WINDIR%\system32\\_up.exe | √ | √ | √ | √ | Event ID<br>Image File Name |
| | | | System | Event ID: 1074<br>Event Msg: system shutdown & restart | x | x | √ | √ | Event ID<br>Event Message |
| | | | Application | Event ID: 1015<br>Event Msg: lsass.exe failed | x | x | √ | √ | Event ID<br>Event Message |
| | Network | Activity | IDS Alert | NETBIOS Unicode share access<br>NETBIOS lsass exploit attempt<br>SHELLCODE detected | √ | √ | √ | √ | Error Message |
| | | Alarm | IDS Alert | Source IP Address: Attacker<br>Destination IP Address: Victim<br>Destination Port: 445 | √ | √ | √ | √ | Source IP Address<br>Destination IP Address<br>Destination Port |

In Security log, the traces data from the security log shows that there is a new process created by system proves by the existence of event id 592. It has installed the FTP server in order to permit the exploit codes down-loaded by other infected computers as shown on the image file name that are *ftp.exe* and *\*_up.exe*.

Meanwhile, Application log shows the traces data of sasser infected machine that indicates the *lsass.exe* application error on the event id 1015 and the event message is *lsass.exe* fail. Subsequently, the infected machine will shutdown and restart again as shown in the *System log* by showing the new process created on event id 1074 and the event message is "*the system process: C:\WINDOWS\system32\lsass.exe terminated unexpectedly with status code 128*". Although these traces data only found in Scenario C, the trace is still significant based on the side effect of sasser worm in which this worm will exploit lsass.exe application, make it crashed and reboot the machine automatically as reported in several antivirus organization such as [16].

The *alert IDS log* shows that there is an activity called lsass exploit attempt on port 445/TCP where the source IP address is the attacker and the destination IP address is the victim. These traces identify that there is a pattern exists on how the sasser worm initiates the communication and exploit the lsass service that used to verifies the validity of user logons to host or server. Lsass generates the process responsible for authenticating users for the Winlogon service. This is performed by using authentication packages. If authentication is successful, *Lsass* generates the user's access token, which is used to launch the initial shell. Other processes that the user initiates then inherit this token. In this case, the worm attempts to obtain the authentication on executing the ftp server to transfer the exploit code to the target.

### 4.2 Worm Trace Pattern Analysis: Attacker Perspective

The attacker's host and network log have been analyzed in order to extract the attacker's data trace in which it will be used as the evidence for the intrusion committed. Based on the analysis, the evidence are found in *personal firewall log, security log, system log, application log* and *ids alert log* in all scenarios as summarized in TABLE 3. The details of the attacker's logs are discussed as following.

The traces data leaved in attacker's *Personal Firewall Log* shown the vulnerable ports that are used by the attacker to open the backdoor in order to exploit the remote shell and transfer the exploit codes on its victims as referred to [17], [18], [16] and [5]. The pattern of the traces data are *445 OPEN TCP, 9996 OPEN TCP, 5554 OPEN-INBOUND TCP* and *OPEN 3\*\*\* TCP*.

The traces data from the security log shows that there is a new process created (Event ID: 592) that shows the sasser worm is activated based on the trace shows on the image file name. The impact of the attack also found from the traces leave on *system log* and *application log*. The traces data shows from *system log* are Event ID: 1074 with the Event Message is system shutdown and restart. On the other hand, the traces data found in *Application log* shows the Event ID: 1015 with Event Message is *lsass.exe failed*. The traces data found in both log is explain the side-effect of the worm is for *LSASS.EXE* to crash, by default such system will reboot after the crash occurs as described in [16].

TABLE 3

SUMMARY OF SASSER TRACES ON ATTACKER'S LOG FOR SCENARIO A, SCENARIO B AND SCENARIO C

(√=TRACES FOUND; x=TRACES NOT FOUND)

| Perspective | Level | Trace Category | Evidence /Log | Trace Pattern | Scenario A | Scenario B | Scenario C | General Attacker Trace Pattern | |
|---|---|---|---|---|---|---|---|---|---|
| | | | | | | | | Trace Discovered | Trace Attributes |
| Attacker | Host | Scan | Personal Firewall | 445 OPEN TCP | √ | √ | √ | √ | Action Protocol Destination Port |
| | | Exploit | Personal Firewall | 9996 OPEN TCP 5554 OPEN-INBOUND TCP 3\*\*\* OPEN TCP | √ | √ | √ | √ | Action Protocol Destination Port |
| | | | Security | Event ID: 592 IFN: ~\sasser.exe | √ | √ | √ | √ | Event ID Image File Name |
| | | Impact | System | Event ID: 1074 Event Msg: system shutdown & restart | √ | √ | √ | √ | Event ID Event Message |
| | | | Application | Event ID: 1015 Event Msg: lsass.exe failed | √ | √ | √ | √ | Event ID Event Message |
| | Network | Activity | IDS Alert | SCANUPnP | √ | √ | √ | √ | Error Message |
| | | Alarm | IDS Alert | Source IP Address: Attacker | √ | √ | √ | √ | Source IP Address |

In the *alert IDS log*, *ScanUPnP* presents the pattern of scanning activity which shows the behavior of traditional worm attack in general and sasser worm attack in specific [19]. Therefore, this trace discovers that the owner of the source IP address is a potential attacker who launched the worm.

### 4.3 Worm Trace Pattern Analysis: Multi-step (Victim/Attacker) Perspective

The multi-step (Attacker/Victim)'s traces data is identified based on the extracted data from the logs at the victim's host and network log in each scenario. The summary of the data traces on the multi-step and network logs for all scenarios are represented in TABLE 4 and the evidences are found in *personal firewall log, security log, system log, application log* and *IDS alert log*. The details of the traces of the multi-step's logs are discussed.

There are two different patterns existing in personal firewall log that discover attacker and victim traces as shown in TABLE 4. The table shows in victim's perspective, the traces data found (*445 OPEN-INBOUND TCP, 9996 OPEN-INBOUND TCP, 5554 OPEN*) indicate that the local host is permitted the FTP service from the remote host. The trace data on *3\*\*\* OPEN-INBOUND TCP* indicates the host allowed to get the malicious codes from the remotes host.

While, from the attacker perspective (*445 OPEN TCP, 9996 OPEN TCP, 5554 OPEN-INBOUND*), the patterns indicate that the local host is opened an outbound session to the remote host which allow the local host transmit the payload (worm codes) to the remote host. The trace data on *3\*\*\* OPEN TCP* indicates the host allowed to communicate on transferring the worm code to the remote host or the target. These communication activities are done by vulnerable ports open exploitation.

TABLE 4
SUMMARY OF SASSER TRACES ON MULTI-STEP (VICTIM/ATTACKER)'S LOG FOR SCENARIO A, SCENARIO B AND SCENARIO C
(√=TRACES FOUND; X=TRACES NOT FOUND)

| Perspective | Level | Trace Category | Evidence /Log | Trace Pattern | Scenario A | Scenario B | Scenario C | General Victim Trace Pattern Trace Discovered | General Victim Trace Pattern Trace Attributes |
|---|---|---|---|---|---|---|---|---|---|
| Victim/ Attacker | Host | Scan | Personal Firewall | VICTIM 445 OPEN-INBOUND TCP  ATTACKER 445 OPEN TCP | √ | √ | √ | √ | Victim/Attacker Action Protocol Destination Port |
| | | Exploit | Personal Firewall | VICTIM 9996 OPEN-INBOUND TCP 5554 OPEN-INBOUND TCP 3\*\*\* OPEN-INBOUND TCP  ATTACKER 9996 OPEN TCP 5554 OPEN TCP 3\*\*\* OPEN TCP | √ | √ | √ | √ | Victim/Attacker Action Protocol Destination Port |
| | | Impact | Security | VICTIM/ATTACKER Event ID: 592 IFN: %WINDIR%\system32\ ftp.exe  Event ID: 592 IFN: %WINDIR%\system32\ *_up.exe | √ | √ | √ | √ | Victim/Attacker Event ID Image File Name |
| | | System | System | VICTIM/ATTACKER Event ID: 1074 Event Msg: system shutdown & restart | x | x | √ | √ | Victim/Attacker Event ID Event Message |
| | | Application | Application | VICTIM/ATTACKER Event ID: 1015 Event Msg: lsass.exe failed | x | x | √ | √ | Victim/Attacker Event ID Event Message |
| | Network | Activity | IDS Alert | VICTIM NETBIOS Unicode share access NETBIOS lsass exploit attempt SHELLCODE detected  ATTACKER SCANUPnP | √ | √ | √ | √ | Victim/Attacker Error Message |
| | | Alarm | IDS Alert | VICTIM Source IP Address: Attacker Destination IP Address: Victim Destination Port: 445  ATTACKER Source IP Address: Attacker | √ | √ | √ | √ | Victim/Attacker Destination IP Address Destination Port  Victim Source IP Address |

Therefore, the traces data found are significant to the multi-step attack (victim/attacker) where this host was infected (act as victim) and as long as the computer was infected with the worm code *(avserve2)*, it (act as attacker) continued to generate traffic to attempt to infect other vulnerable computers [17].

The traces data in *security log* in TABLE 4 shows that there is a new process created (*Event ID: 592*) by system which initiates the *FTP* service in all scenarios. This service is used to receive and sent the sasser worm code *(\*_up.exe)* and execute the sasser worm code *(\*_up.exe or avserver2.exe: if the host is rebooted)* remotely. This trace pattern indicates that this host is became a victim of worm attack that received the sasser code *(\*_up.exe)* and became an attacker to another target by executing the sasser code *(\*_up.exe)* as shown on the image file name. It identify that this host was infected previously and automatically attempt to transfer the worm code by generating traffic to exploit other vulnerable computers.

*System log* shows the traces data of the Sasser-infected machine stops in Scenario C by showing the new process created on event id 1074 that indicates the system shutdown and restart. This data traces can be support by the traces data found in *Application log* that showing the new process created on event id 1015 that indicates the lsass.exe application is failed. These patterns are signifi-cant with the victim pattern in which if the host is infected by sasser worm, the lsass.exe application is failed and crashed that force the windows restart.

TABLE 4 also describes there are traces found in the *alert IDS log* are *NETBIOS Unicode share access*, *NETBIOS lsass exploit attempt* and *SHELLCODE detected*, and *SCANUPnP* activities for victim and attacker respectively. The *NETBIOS Unicode share access*, *NETBIOS lsass exploit attempt* and *SHELLCODE detected* activities trace indicates that there is a pattern exists on how the sasser worm initiates the client to share and exploit the lsass application. These traces identify the source IP address is the attacker and the destination IP address is the victim. On the other hand, the *SCANUPnP* trace proves that there is scanning activity on the vulnerable open port. This trace indicates that the source IP address is the attacker who activated the worm. Both traces found in TABLE 4 are significant to the pattern that found in victim and attacker.

The summary of the analysis as shown in TABLE 2, TABLE 3 and TABLE 4 identified findings on the significant attributes from victim, attacker and multi-step traces data. Hence, these findings are further use to form the proposed general worm trace pattern.

## 5 PROPOSED GENERAL WORM TRACE PATTERN

This research proposed a general worm trace pattern based on victim, attacker and multi-step perspective. The details are described in this section as the following.

### 5.1 General Victim's Trace Pattern

Victim's trace pattern can be used as a guide during digital forensic investigation in order to provide a precise hypothesis on how the intrusion happened such as how the victim attacked by the potential attacker. In this research, a general Sasser victim's trace pattern is established as depicted in Fig. 4 based on the findings from TABLE 2.

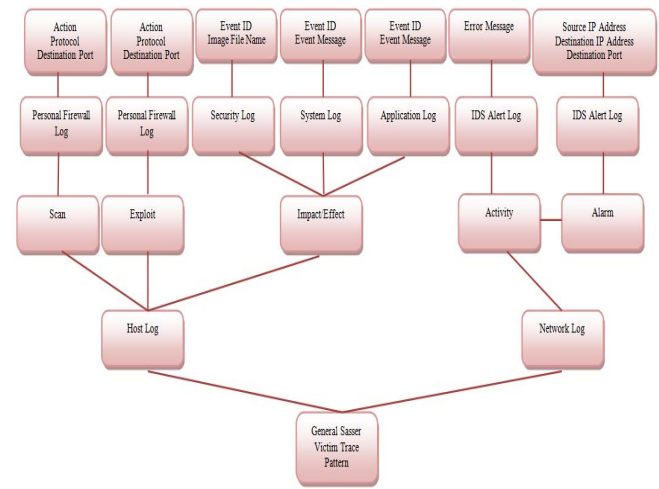

Fig.4 Proposed General Sasser Victim's Trace Pattern

From the traces found as shown in Fig. 4, from personal firewall log and security log indicate the worm trace pattern at the victim's host used vulnerable open port with specific protocol to permit the scanning and transmitting the exploit codes from remote host which launch the windows shell to initiate worm code download. In this research, the findings found that Sasser trace pattern at the victim's host identified in the personal firewall log shows this worm is used port *445 TCP, 9996 TCP, 5554 TCP* and *3\*\*\* TCP* to complete its activities on scanning and exploiting the target host. These activities had been confirmed by traces found in security log about the impact from the attack that shows new process created (Event ID: 592) on FTP service and *\*_up.exe* as shown in image file name. These traces indicate how sasser worm transfer the exploit code (*\*_up.exe*) to the target host.

The traces also found from *system log*, *application log* and *IDS alert log*. In system log and application log, the traces found from the attributes *Event ID* and *Event Message* show the impact of the attack. Meanwhile, the intrusion activity and alarm had been traced from the IDS alert logs based on the attribute: *error message*, source IP address, destination IP address and destination port. The trace of source IP address and destination IP address in the IDS alert logs identified the victim and the attacker respectively.

In this research, the traces found indicate the impact of the sasser worm intrusion that shows trace of the lsass.exe application failed found in *application log* that caused system crashed which initiate the system shutdown. This impact has been proved as the trace of system shutdown and restart is found in the *application log*. The traces found from the *IDS alert log* also supports all the traces found on the host log such as the *lsass exploit attempt* which explain exploiting activities done by sasser worm.

## 5.2 General Attacker's Trace Pattern

Attacker's trace pattern is useful to guide forensic investigators in searching the evidence and provide a structured method for obtaining and representing relevant digital forensic information. This pattern provides a systematic description of the attack goals and strategies for tracing the attack.

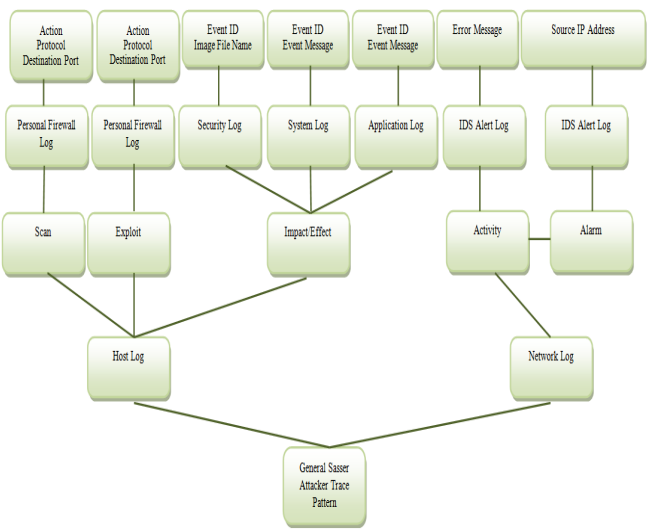

Fig. 5 Proposed General Sasser Attacker's Trace Pattern

The overall Sasser attacker's trace pattern that found in various logs as depicted in Fig. 5 is derived based on the findings in TABLE 3. The traces indicate the sasser worm pattern at the attacker's host used port *445 TCP* to allow the local host scan and transmit exploit codes to the remote host which launch the windows shell to initiate worm code download used port *9996 TCP*. Then it launched the *FTP* client service using port *5554 TCP* and open port *3\*\*\* TCP* to permit the client (remote host) download the worm code from the local host. The activity of *SCANUPnP* trace that explains the scanning activities done by sasser worm also existed in the network log that supports all the traces found on the host log.

## 5.3 General Multi-step (Victim/Attacker)'s Trace Pattern

Multi-step's trace pattern is used as a guide for forensic investigators to reveal and prove the true attacker or victim. This trace pattern is a combination of victim's and attacker's trace pattern in which the traces data is extracted from a log for the same host that divided into primary and secondary evidence.

The overall traces data on multi-step at the host's logs from victim/attacker perspective illustrated in Fig. 6 indicate that the sasser worm used port *445 TCP* to permit the scanning activity and it is supported by the traces found in network logs that show *SCANUPnP* activities.

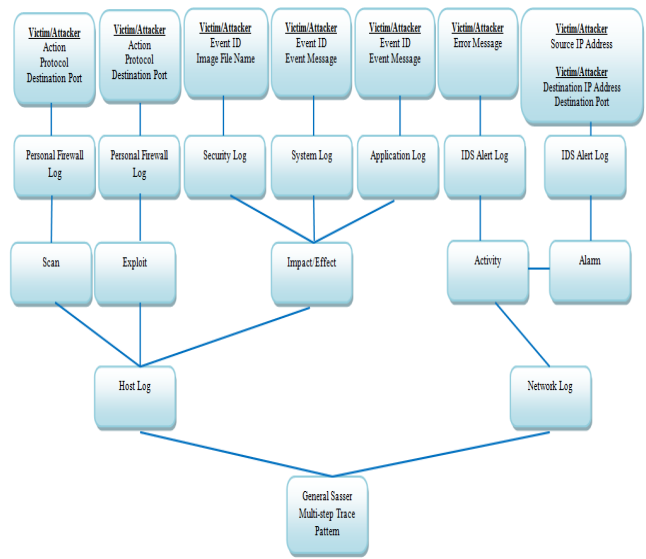

Fig. 6 Proposed General Multi-step (Victim/Attacker)'s Trace Pattern

The worm then transmit exploit codes from remote host which launch the windows shell to initiate downloading the worm code using port *9996 TCP* and it launched the FTP client service on port *5554 TCP* and open port *3\*\*\* TCP* to download/upload the exploit codes. This worm activity is shown by the traces found in network logs that confirm the existence of *lsass exploit attempt* activities. Once the host is infected (act as victim), it's (act as attacker) then generate traffic; attempt to infect other vulnerable hosts.

The source IP address from host log indicates that the remote host is the attacker and the destination IP address which is the local host is the victim. Hence, multi-step (victim/attacker) trace pattern could identify the true victim or attacker.

## 6  CONCLUSIONS AND FUTURE WORKS

Trace pattern of an intrusion in an intrusion scenario is constructed by analyzing heterogeneous logs from diverse devices in victim, attacker and victim/attacker (multi-step) perspectives. These trace patterns offer a systematic description of the impact of the intrusion, the intrusion goals and intrusion strategies for tracing the attack in order to create a precise hypothesis about the intrusion on revealing the attacker or victim. For example, personal firewall logs provide information on how the attacker entered the network and how the exploits were performed; meanwhile event logging such as security log, system log and application log enables network administrators to collect important information such as date, time and result of each action during the setup and execution of an intrusion. Therefore, the propose victim, attacker and multi-step (victim/attacker) trace patterns in this paper can be extended into research areas in alert correlation and computer forensic investigation.

## ACKNOWLEDGMENT

We thank to Universiti Teknikal Malaysia Melaka for the Short Grant Funding (PJP/2010/FTMK (10D) S693) for this research project.